\documentclass[twocolumn, prl, showpacs, floatfix]{revtex4}

\usepackage{amsmath}
\usepackage{graphicx}

\begin{document}

\title{Phase Transition in a Self-repairing Random Network}
\author{A.~S.~Ioselevich, D.~S.~Lyubshin}
\affiliation{Landau Institute for Theoretical Physics, Russian
 Academy of Sciences, Kosygina str.2, 117940  Moscow, Russia}
\date{\today}
\begin{abstract}
We consider a network, bonds of which are being sequentially
removed; that is done at random, but conditioned on the system
remaining connected (Self-Repairing Bond Percolation SRBP). This
model is the simplest representative of a class of random systems
for which forming of isolated clusters is forbidden. It
qualitatively describes the process of fabrication of artificial
porous materials and degradation of strained polymers. We find  a
phase transition at a finite concentration of bonds $p=p_c$, at
which the backbone of the system vanishes; for all $p<p_c$ the
network is a dense fractal.
\end{abstract}
\maketitle

Properties of networks, bonds (or nodes) of which are being
removed randomly, have been intensively studied during past 50
years. In the standard formulation of the problem the bonds (or
sites) are removed totally at random, and the percolation
transition takes place at a certain concentration $p_c$ of
remaining bonds (or concentration $x_c$ of remaining sites). For
$p>p_c$ there exists an ``infinite'' cluster that contains a finite
fraction of all bonds in the system and spans homogeneously through
the entire network. At $p<p_c$ the infinite cluster does not exist
and only finite ones are present. The percolation phase transition
and corresponding critical phenomena are well studied (see, e.g.,
\cite{stauffer-aharony, bunde-havlin}).

However, there are many physical systems which cannot be described
by the standard percolation theory. In particular, there are
important cases when finite clusters can not appear at all.  As an
example, consider a technologically important process of pore
forming (i.e., fabrication of a porous material; see, e.g.,
\cite{book}). It can be viewed as gradual removal of grains of a
pore-former (carbon, which can be burned out, or a soluble
polymer) from a mixture of the pore-former with grains of a
matrix material (a metal). Due to mechanical instability of finite
clusters, they immediately fall down onto the surrounding matrix
and stick to it.  Thus at any stage of the process the remaining grains
form a single ``infinite cluster'' and the percolation transition is
impossible.  The  properties of this single cluster
are nontrivial: it appears that there is a topological phase
transition at a finite concentration of remaining grains $x_c$,
below which the system becomes ``cracked'' and its mechanical and
conducting properties degrade catastrophically.

So far we were not able to demonstrate the existence of the above
phase transition by analytic methods, our qualitative arguments
and results of numerical studies will be published elsewhere
\cite{single-cluster}. In the present paper we introduce a
simplified model, which allows rigorous analysis. The model is
very similar to the standard bond percolation: starting from a
full lattice, at each step one of the remaining bonds is randomly
chosen for removal.  But after its removal the system is checked
for the existence of finite clusters: if such ones are
present, then the removed bond is restored (i.e., the last removal
is cancelled) and the process goes on to the next step. It seems
natural to call this model a ``self-repairing bond percolation''
(SRBP).

Apart from being relevant to pore-forming, the SRBP model may also
be viewed as a model for polymer degradation (see, e.g.,
\cite{polymer}). Consider a random network consisting of
irregularly cross-linked polymer chains. Suppose that this system
is subjected to random external perturbation (e.g., UV-radiation)
that can destroy the cross-links.  The radiation damage may
be repairable: attraction between
individual chains tends to reestablish the damaged link. However,
sometimes that appears to be impossible, since internal strains in the chains
may drive the two chains apart as soon as the link
between them is damaged. Thus it seems reasonable to assume that
all strained links are vulnerable to radiational damage, while
unstrained ones are ``immune'' to it.  Of course, finding out
which links in a random network are strained and which are not is
a formidable task. But in any case the links which are the only bridges
connecting otherwise isolated clusters are
never strained. These links will be repaired after
possible removal, in accordance with the definition of the SRBP model.

Let the fraction of the remaining bonds be $p$. We will be
interested in average properties of the system as function of $p$.
In particular, we will study the conductivity $\sigma(p)$ and
``the minimal chemical path'' $\ell (R,p)$, the latter being an
ensemble-averaged length of the shortest path going via bonds and
connecting two points separated by euclidean distance $R$.

The first obvious observation about the SRBP model is that there
exists some minimal possible $p=p_{\rm tree}\equiv \frac{2}{z}$
($z$ being the coordination number of the lattice) at which the
process of bond removal stops: for a connected graph one
necessarily has $p\geq p_{\rm tree}$.  At $p=p_{\rm tree}$ the
remaining bonds constitute a spanning tree (ST), a connected graph
with no cycles and all lattice sites as vertices.  As it is
well known, the probability of generating a given ST at the end of
our process is related to the Minimal Spanning Tree (MST) problem
(see, e.g., \cite{dobrin-duxbury}). In particular, the minimal
chemical path is fractal: $\ell (R,p=p_{\rm tree})\propto
R^{d_{\rm min}^{(\rm MST)}}$ with $d_{\rm min}^{(\rm MST)}>1$.
Obviously, for a tree one has $\sigma(p=p_{\rm tree})=0$.

We will show that actually the minimal chemical path is fractal
and the specific conductivity of the system is zero not only at
$p=p_{\rm tree}$, but also within a finite interval $p_{\rm
tree}\leq p\leq p_{\rm c}$. The corresponding phase we will call
the ``tree-like phase'', in contrast to the ``solid phase''
existing at $p_{\rm c}\leq p<1 $.

\begin{figure}
\includegraphics[scale=0.33]{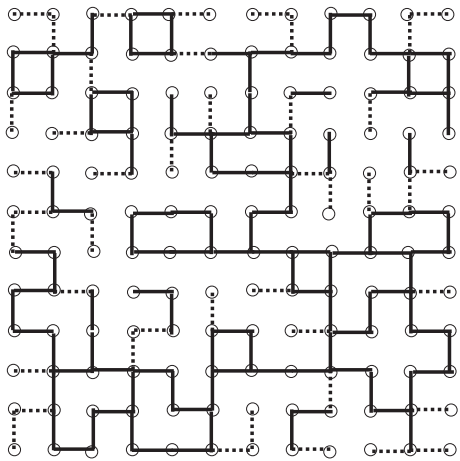}
\caption{A typical configuration of black ang grey bonds at
 the critical concentration for 2D square lattice.
 }
 \label{black-and-grey}
\end{figure}

To prove the above statement, we use a mapping to the standard
percolation. Suppose that initially all the bonds of the system
are black. If at some step a removal of a certain bond must be
cancelled, we restore the bond but change its color to grey. Then
for any fraction $p$ of remaining bonds we have fractions
$b=b(p)$ of black and $g=p-b(p)$ of grey bonds remaining, where
$b(p)$ is a certain monotonically increasing function with the
following asymptotics:
\begin{equation}
 b(p)\approx
 \left\{
  \begin{aligned}
   p & \quad    \mbox{for $1-p\ll 1$},\\
   0 &  \quad   \mbox{for $p\to p_{\rm tree}$}.
  \end{aligned}
 \right.
\label{black}
\end{equation}
Clearly, a grey bond may never be removed, and at $p=p_{\rm tree}$
all bonds are either removed or grey.  It is easy to see that the
backbone of the entire (black and grey) network coincides with the
backbone of the black subsystem.  Indeed, no grey bond can belong
to backbone: since its removal produces an isolated finite
cluster, such a bond belongs to a dangling end. Note
that black bonds are removed {\em totally at random}, hence the
behavior of the black subsystem is identical to that of the
standard bond-percolation system.  In particular, the backbone
vanishes at the percolation point, where $b=p_{\rm perc}$. It
follows that there exists a critical concentration of bonds $p_c$
such that for $p<p_c$ the remaining bonds all belong to one
infinite cluster (which has finite density), while this cluster
{\em has no backbone}. The critical concentration is determined by
the condition
\begin{equation}
 b(p_c)=p_{\rm perc},
 \label{threshold}
\end{equation}
where $p_{\rm perc}$ is the percolation threshold for the standard
bond percolation problem on the same lattice.  On the other hand,
the number of grey bonds in the system is equal to the number of
finite black clusters: $g=n_{\rm cl}$ (see Fig.
\ref{black-and-grey}). Thus the critical concentration $p_c$ can
be expressed solely through the characteristics of the standard
percolation problem:
\begin{equation}
 p_c=p_{\rm perc}+n_{\rm cl}^*,
 \label{threshold1}
\end{equation}
where $n_{\rm cl}^*$ is the number of finite clusters (per one
bond of the initial lattice) at the critical point. The latter is
known for many lattices; in
particular, for the square lattice $n_{\rm cl}^*=(3\sqrt{3}-5)/4$
(see \cite{temperley-lieb}) and $p_{\rm perc}=1/2$, so
\begin{equation}
 p_c=\frac{3\sqrt{3}-3}{4}\approx 0.54904.
 \label{threshold2}
\end{equation}

\begin{figure}
\includegraphics[scale=0.33]{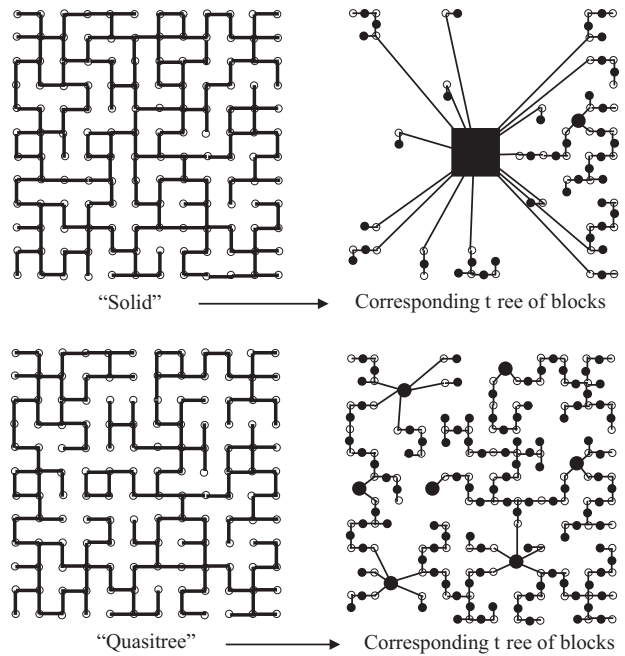}
\caption{
 Typical configurations of bonds in the solid and tree-like phases
 (we do not distinguish black and grey bonds here),
 together with corresponding block graphs. Black square---the
 infinite block (the backbone); large solid circles---finite
 nontrivial blocks; small solid circles---trivial blocks;
 small open circles ---articulation points.
}
\label{blocks}
\end{figure}

In scaling theory of percolation the relations between different
critical exponents are normally derived from considerations
involving distribution function of finite clusters (see, e.g.,
\cite{bunde-havlin}). In our model finite clusters do not exist
whatsoever, but fortunately one can introduce blocks---alternative
objects, which to some extent play the role of finite clusters and
make it possible to develop the scaling theory. By definition, a
{\em block} is a maximal subgraph that cannot be disconnected by
deletion of a single vertex \cite{harary}. It is not difficult to
show that either a block consists of a single bond (and its two
ends), or any two bonds belonging to a block lie on a common
cycle.  Two distinct blocks may have at most one point in common;
such a point is called an {\em articulation point}, and its
deletion necessarily disconnects the system.  Given a network, one
can form a graph with  blocks and articulation points of the
network as vertices, with two vertices connected if they
correspond to an articulation point and a block that contains it.
Such a {\em block graph} is always a tree; an example is shown in
Fig. \ref{blocks}.

The backbone that exists in the solid phase constitutes the only
infinite block in the system; it has finite density and contains
infinite cycles. The backbone is linked to an infinite number of
branches---dangling ends, each dangling end being a finite tree of
finite blocks.

In the tree-like phase the infinite block collapses, so that there
are only finite cycles in this phase.  The corresponding bond
configurations we will call quasitrees.

In the solid phase, the backbone becomes more loose as $p$
approaches $p_c$.  The fraction of bonds belonging to it tends to
zero:
\begin{eqnarray}
 P_B(p)\propto (p-p_c)^{\beta_B},
 \label{backbone}
\end{eqnarray}
where $\beta_B$ is the index of the backbone density for the
standard percolation (in particular, $\beta_B\approx 0.48$ for
$d=2$, see, e.g., \cite{bunde-havlin} and \cite{grassberger}).

The total number of dangling ends decreases as $p\to p_c$ from
above, while the number of blocks in each dangling end and the
number of bonds in a typical block increase and diverge as $p\to
p_c$.  It is convenient to introduce the distribution function of
finite blocks consisting of $l$ bonds:
\begin{eqnarray}
 n_l(p)\sim l^{-\tau}f\left[l(p-p_c)^{1/\sigma}\right],
 \label{distribution}
\end{eqnarray}
where $f(x)$ is a universal function that decays exponentially at
$x\gg 1$.  Application of standard scaling arguments
\cite{bunde-havlin} to blocks lead to the following relations
between the critical exponents:
\begin{eqnarray}
 \nu=\frac{\tau -1}{\sigma d},\qquad \xi(p)\sim (p-p_c)^{-\nu}
 \label{correlation}
\end{eqnarray}
The length $\xi(p)$ characterizes correlations within the
backbone, and since the backbone for the SRBP model is the same as
for standard percolation, we conclude that the exponent $\nu$ for
the SRBP model coincides with that for standard percolation.

The exponent $\gamma$ that characterizes the behavior of the mean
size $S(p)$ of finite blocks near $p_c$ is
\begin{eqnarray}
 \gamma=\frac{3-\tau}{\sigma},\qquad
 S(p)\equiv\frac{\sum_ll^2n_l(p)} {p-P_B(p)}\propto (p-p_c)^{-\gamma}.
 \label{size}
\end{eqnarray}
Finally,
\begin{eqnarray}
 \beta_B=\frac{\tau -2}{\sigma},\qquad
 d_B=\frac{d}{\tau -1},
 \label{scaling}
\end{eqnarray}
where $d_B$ is the fractal dimension of the backbone at $p=p_c$.

Since the conduction process involves only the backbone, the
conductivity of the SRBP model is identical to that of the
standard percolation,
\begin{eqnarray}
 \sigma_{\rm SRBP}(p)\equiv\sigma_{\rm perc}[b(p)]\propto (p-p_c)^{\mu},
 \label{cond}
\end{eqnarray}
hence the corresponding critical exponent $\mu$ is the standard
one.  For the minimal path length in the solid phase one has
\begin{equation}
 \begin{gathered}
  \ell (R)=\frac{R}{v(p)},\\
  v(p)\sim\xi(p)^{1-d_{\rm min}^{(\rm perc)}}\sim
  (p-p_c)^{-\nu+\nu d_{\rm min}^{(\rm perc)}},
 \end{gathered}
 \label{path}
\end{equation}
where $d_{\rm min}^{(\rm perc)}$ is the graph dimension for the
infinite cluster at the critical point for the standard
percolation problem.  As usual, the formula (\ref{path}) is valid
only for $R\gg\xi(p)$; in the opposite case $R\ll\xi(p)$ it should
be substituted by the critical law
\begin{eqnarray}
 \ell (R)\propto R^{d_{\rm min}^{(\rm perc)}}.
 \label{pathx}
\end{eqnarray}
For $R\sim\xi(p)$ the expressions (\ref{path}) and (\ref{pathx})
match.

\begin{figure}
\includegraphics[scale=0.33]{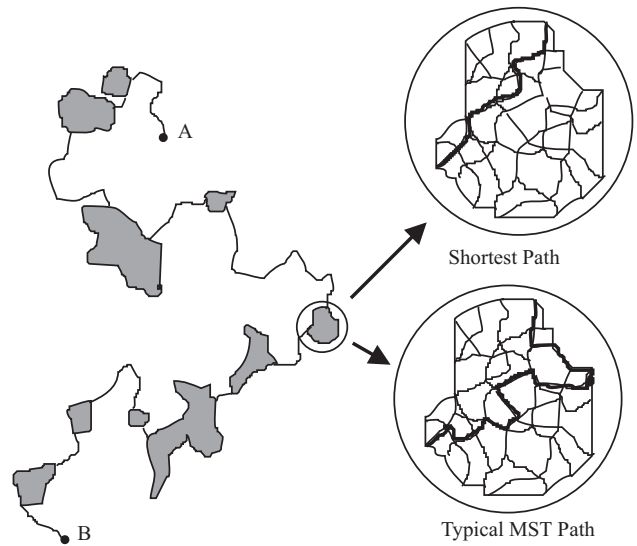}
\caption{The path, connecting points $A$ and $B$ on the block graph;
 nontrivial blocks are shown as grey islands.
 The internal structure of a typical block is shown in inserts:
 a typical MST-path, traversing a block (lower panel);
 the shortest path across the block (upper panel).}
\label{block-path}
\end{figure}

At $p=p_{\rm tree}$ our system is reduced to the MST ensemble.
In high dimensions $d>d_c$ a minimal spanning tree on an
{\em infinite} lattice may in fact have many components.
It is believed (see \cite{newman-stein}) that $d_c=8$, and in
$d<8$ dimensions for almost all trees of the MST ensemble there
exists a unique finite path ${\cal P}(A, B)$ connecting any two
given sites $A$ and $B$. This path is a certain non self-intersecting
random walk with fractal dimension $d_f=d_{\rm min}$. While this
dimension is known exactly for the 2D uniform spanning tree (UST)
ensemble, only numerical estimates are available for the MST case:
$d_{\rm min}\approx 1.22$ for $d=2$,
$d_{\rm min}\approx 1.42$ for $d=3$,
and $d_{\rm min}\approx 1.59$ for $d=4$
(see \cite{numerical dmin}). Below we demonstrate that
the graph dimension is the same {\em throughout the entire
tree-like phase}:
\begin{eqnarray}
d_{\rm min}(p)=d_{\rm min}^{(\rm MST)}.
\label{path0}
\end{eqnarray}
More precisely,
\begin{equation}
 \begin{gathered}
  \ell (R,p)=\frac{R^{d_{\rm min}^{\rm(MST)}}}{v(p)}, \\
  v(p)\sim\xi(p)^{d_{\rm min}^{\rm(MST)}-d_{\rm min}^{\rm
 (perc)}}\sim (p_c-p)^{-\nu d_{\rm min}^{\rm(MST)}+\nu d_{\rm min}^{\rm (perc)}},
 \end{gathered}
 \label{path1}
\end{equation}

\begin{figure}
\includegraphics[scale=0.7]{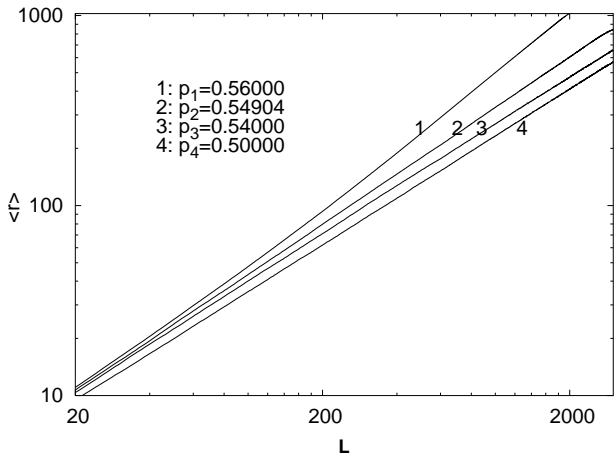}
\caption{Root mean square euclidean displacement vs.~chemical
distance for different values of $p$, obtained via averaging over
$1600$ realizations of a $2048\times 2048$ lattice with periodic
boundary conditions. Asymptotic slopes correspond to $d_{\rm
min}(p_c)=1.13(1)$ and $d_{\rm min}(p_c-0.01=0.54)=d_{\rm
min}(p_{\rm tree})=1.22(1)$ in agreement with known results.}
\label{numerical}
\end{figure}

Consider a certain quasitree ${\cal Q}$ and the set of trees
${\cal T}_i^{({\cal Q})}$ which can originate from ${\cal Q}$ in
the course of further destruction of bonds. Obviously, ${\cal Q}$
is a union of all ${\cal T}_i^{({\cal Q})}$:
\begin{eqnarray}
 {\cal Q}=\cup_i{\cal T}_i^{({\cal Q})}.
 \label{bundle1}
\end{eqnarray}
Now  we introduce a graph
\begin{eqnarray}
 \overline{\cal P}(A,B)^{({\cal Q})}=\cup_i{\cal P}_i(A, B)^{({\cal Q})},
 \label{bundle2}
\end{eqnarray}
which is the union of all paths ${\cal P}_i^{({\cal Q})}$ leading
from $A$ to $B$ in all trees ${\cal T}_i^{({\cal Q})}$. It is easy
to show that $\overline{\cal P}(A, B)^{({\cal Q})}$ is precisely
the path leading from $A$ to $B$ {\em in the block graph} (see
Fig. \ref{block-path}). The minimal (over the entire quasitree)
path ${\cal P}(A, B)^{({\cal Q})}$ leading from $A$ to $B$ is,
obviously,  the minimal path over the graph $\overline{\cal P}(A,
B)^{({\cal Q})}$.  On nontrivial (containing more than one bond)
blocks the path ${\cal P}(A,B)^{({\cal Q})}$ is the shortest path
that crosses the block; it may be considerably shorter than any
individual MST-path.  This consideration enables one to estimate
the typical ratio of lengths for a piece of the minimal path
${\cal P}(A,B)^{({\cal Q})}$ and the corresponding piece of the
MST minimal path. We make such an estimate for the case when
$p<p_c$ but $p_c-p\ll 1$ (i.e., for the vicinity of the phase
transition). Having in mind that the typical block size is
$\xi(p)\gg 1$, for a typical length MST-path crossing such a block
we get $\ell_{\rm MST}(\xi)\sim\xi^{d_{\rm min}^{\rm(MST)}}$,
while the {\em shortest} path traversing the block is the same as
for the critical percolation: $\ell_{\rm
short}(\xi)\sim\xi^{d_{\rm min}^{\rm(perc)}}$. As a result, we
arrive at the estimate (\ref{path1}), which  matches with
(\ref{pathx}) for the critical case $R\sim \xi$.

Since the above consideration is not quite rigorous, we have also
undertaken numerical evaluation of $d_{\rm min}(p)$ in order to
check the identity (\ref{path0}). Simulations did not show any
variation of $d_{\rm min}$ with $p$ (see Fig. \ref{numerical}).

In conclusion, we have demonstrated, both numerically and
analytically, that the Self-Repairing Bond Percolation model
undergoes a topological phase transition at a certain concentration
$p_c$ of remaining bonds. In the tree-like phase (for $p<p_c$) the
network, although being fully connected, has {\em no backbone} and
hence zero conductivity.  The corresponding graphs of bonds
are ``quasitrees'': they contain only finite cycles (even for the
infinite lattice). The properties of the statistical ensemble of
quasitrees are similar to those of the Minimal Spanning Trees
ensemble.

\end{document}